\documentclass[12pt]{article}
\usepackage{graphicx}
\usepackage{url}

\newcommand\pubnumber{SLAC--PUB-16315}
\newcommand\pubdate{June, 2015}

\def\SLAC{SLAC,
    Stanford University, Menlo Park, California 94025 USA}
\def\doeack{\footnote{Work supported by the US Department of Energy,
                     contract DE--AC02--76SF00515.}}

\def\Title#1{\begin{center} {\Large #1 } \end{center}}
\def\Author#1{\begin{center}{ \sc #1} \end{center}}
\def\Address#1{\begin{center}{ \it #1} \end{center}}

\def\submit#1{\begin{center}Submitted to {\sl #1} \end{center}}
\newcommand\pubblock{\rightline{\begin{tabular}{l} \pubnumber\\
         \pubdate \end{tabular}}}
\newenvironment{Abstract}{\begin{quotation} \begin{center}
                       ABSTRACT
     \end{center}\bigskip  }{\end{quotation}}

\def\submit#1{\begin{center}Submitted to {\sl #1} \end{center}}
\def\Acknowledgements{\bigskip  \bigskip \begin{center} \begin{large}
             \bf ACKNOWLEDGEMENTS \end{large}\end{center}}

\def\beq{\begin{equation}}
\def\eeq#1{\label{#1}\end{equation}}
\def\eeqn{\end{equation}}

\newenvironment{Eqnarray}%
   {\arraycolsep 0.14em\begin{eqnarray}}{\end{eqnarray}}
\def\beqa{\begin{Eqnarray}}
\def\eeqa#1{\label{#1}\end{Eqnarray}}
\def\eeqan{\end{Eqnarray}}

\def\leqn#1{(\ref{#1})}

\let\bar=\overbar

\def\VEV#1{\left\langle{ #1} \right\rangle}
\def\bra#1{\left\langle{ #1} \right|}
\def\ket#1{\left| {#1} \right\rangle}

\def\lsim{\mathrel{\raise.3ex\hbox{$<$\kern-.75em\lower1ex\hbox{$\sim$}}}}
\def\gsim{\mathrel{\raise.3ex\hbox{$>$\kern-.75em\lower1ex\hbox{$\sim$}}}}

\def\M{{\cal M}}

\def\half{\frac{1}{2}}

\def\tthird{\frac{2}{3}}

\def\del{\partial}
\def\Dslash{\not{\hbox{\kern-4pt $D$}}}
\def\dslash{\not{\hbox{\kern-2pt $\del$}}}

\def\ee{e^+e^-}

\def\msb{{\bar{\scriptsize M \kern -1pt S}}}

\def\drb{{\bar{\scriptsize D \kern -1pt R}}}

\def\eps{\epsilon}

\def\s#1{\widetilde{#1}}

\makeatletter
\def\section{\@startsection{section}{0}{\z@}{5.5ex plus .5ex minus
 1.5ex}{2.3ex plus .2ex}{\large\bf}}
\def\subsection{\@startsection{subsection}{1}{\z@}{3.5ex plus .5ex minus
 1.5ex}{1.3ex plus .2ex}{\normalsize\bf}}
\def\subsubsection{\@startsection{subsubsection}{2}{\z@}{-3.5ex plus
-1ex minus  -.2ex}{2.3ex plus .2ex}{\normalsize\sl}}

\renewcommand{\@makecaption}[2]{%
   \vskip 10pt
   \setbox\@tempboxa\hbox{\small #1: #2}
   \ifdim \wd\@tempboxa >\hsize     
       \small #1: #2\par          
     \else                        
       \hbox to\hsize{\hfil\box\@tempboxa\hfil}
   \fi}

 \def\citenum#1{{\def\@cite##1##2{##1}\cite{#1}}}
 
\newcount\@tempcntc
\def\@citex[#1]#2{\if@filesw\immediate\write\@auxout{\string\citation{#2}}\fi
  \@tempcnta\z@\@tempcntb\m@ne\def\@citea{}\@cite{\@for\@citeb:=#2\do
    {\@ifundefined
       {b@\@citeb}{\@citeo\@tempcntb\m@ne\@citea\def\@citea{,}{\bf ?}\@warning
       {Citation `\@citeb' on page \thepage \space undefined}}%
    {\setbox\z@\hbox{\global\@tempcntc0\csname b@\@citeb\endcsname\relax}%
     \ifnum\@tempcntc=\z@ \@citeo\@tempcntb\m@ne
       \@citea\def\@citea{,}\hbox{\csname b@\@citeb\endcsname}%
     \else
      \advance\@tempcntb\@ne
      \ifnum\@tempcntb=\@tempcntc
      \else\advance\@tempcntb\m@ne\@citeo
      \@tempcnta\@tempcntc\@tempcntb\@tempcntc\fi\fi}}\@citeo}{#1}}
\def\@citeo{\ifnum\@tempcnta>\@tempcntb\else\@citea\def\@citea{,}%
  \ifnum\@tempcnta=\@tempcntb\the\@tempcnta\else
  {\advance\@tempcnta\@ne\ifnum\@tempcnta=\@tempcntb \else\def\@citea{--}\fi
    \advance\@tempcnta\m@ne\the\@tempcnta\@citea\the\@tempcntb}\fi\fi}
\makeatother

\def\La{{\cal L}}

\begin{document}
\begin{titlepage}
\pubblock

\vfill
\Title{On the Trail of the Higgs Boson}
\vfill
\Author{Michael E. Peskin\doeack}
\Address{\SLAC}
\vfill
\begin{Abstract}
I review theoretical issues associated with the Higgs boson and the
mystery  of  spontaneous breaking of the electroweak gauge symmetry.   This essay is intended as
an introduction to 
the special issue of {\sl Annalen der Physik},   ``Particle Physics after the Higgs''.
\end{Abstract}
\vfill
\submit{Annalen der Physik}
\vfill

\newpage
\tableofcontents
\end{titlepage}

\def\thefootnote{\fnsymbol{footnote}}
\setcounter{footnote}{0}

\section{Introduction}

The discovery of the Higgs boson in July 2012~\cite{ATLAS,CMS}  brought particle physics
to a significant milestone.  Since the 1970's, we have had a
``Standard Model'' of the strong, weak, and electromagnetic
interactions that has seemed to describe all elementary particle
phenomena observed at high-energy accelerators.   Over forty years,
this model has passed a series of increasingly stringent tests.   As the
parameters of the model became better defined and its predictions
tested more incisively, points of
disagreement between theory and experiment have faded away.  Now the
last elementary particle predicted by this model has been observed.

This discovery has cast the field of particle physics into a high
state of tension.  It is possible that our  understanding of nature's particles and
forces is complete, at least for the foreseeable future.  It is
equally well possible that the trail we have been following will veer
off to reveal a completely new set of particles and interactions.
The arguments on both sides rely on properties of the Higgs
boson. 

Thus, to introduce a volume on future experiments and facilities in
high-energy physics, it is valuable to review what is known about the
Higgs boson  and what
is expected from it.   
To what extent does the Standard Model provide a beautiful and
simple theory that solves the problem for which the Higgs boson was
invented?   To what extent does this solution still leave mysteries?
To what extent is this theory inadequate and in need of replacement?
I will address these questions in this review.

There are many arguments outside the domain of the Higgs boson that
the Standard Model is incomplete as a description of nature.   The
Standard Model does not include gravity.   The Standard Model does not
explain the small size of the cosmological constant or provide an
alternative explanation for the accelerating expansion of the
universe.  The Standard Model does not have a place for the dark
matter that makes up 85\% of the mass in the unverse.   The Standard
Model does not have a place for neutrino masses (though these are
readily accounted for by adding right-handed neutrinos).   The
Standard Model does not explain the preponderance of matter over
antimatter in the universe.  Solutions to these questions might imply
new phenomena that will be observed in the near future.  The review
\cite{Snowmass} describes these possibilities.  On the other hand, all of these
questions have possible explanations for which the observation of new
physics is far out of reach. In this review, I will put all of
these issues aside and speak only about the implications that we can
draw from the physics of the Higgs boson.

\section{Why do we need the Higgs boson?}

To discuss whether the Standard Model (SM) gives an adequate theory of
the Higgs boson, we first need to review the reasons that the Higgs
boson is needed.  Our long wait for the Higgs discovery
encouraged the idea that this particle might not actually  exist.    It was
always possible that the Higgs boson could have been very heavy and
difficult to discover.  However, we could not have lived without
it.    The Higgs field, the field for which the boson is a quantum
excitation, fills two essential  purposes in the SM. It is
necessary to provide the masses of quarks and leptons, and it is
necessary to provide the masses of the $W$ and $Z$ bosons.  In this
section, I will review the evidence for these statements.

\subsection{Electroweak quantum numbers}

It is natural to think that there is no difficulty in generating
masses for elementary particles.   Particles should just {\it have} mass.
However, in relativistic quantum field theory, this point of view is 
naive and incorrect.  

To build a wave equation for a relativistic field, we start from 
fields that transform under the Poincar\'e group according to the basic
irreducible representations.  The simplest such representations, and
the ones that serve as the building-blocks for all others, are
representations for massless fields.  The wave equations for these
fields describe two particle states---a massless particle with
helicity $J$ (an integer or half-integer) and a massless antiparticle 
with helicity $-J$~\cite{SWeinberg}. 

A massles particle always
travels at the speed of light. We can never stop it
and manipulate it at rest. This is the reason that the restriction to
a single helicity state is consistent.  If a particle has mass,  
we can bring it to rest by a Lorentz transformation and then  turn
it in an arbitrary direction.  This will rotate the state of
helicity $J$ into  any linear combination of the  $(2J+1)$ states of a
spin $J$ representatioin of the rotation group.  If the particle is not its own antiparticle, another
$(2J+1)$ antiparticle states  are also required.  Thus, special  relativity places an essential
barrier to  promoting a zero-mass particle to nonzero mass.  
This promotion can only be done if we can identify the $(2J+1)$ elementary
Poincar\'e representations that will form the massive particle, and if
we can  mix
these states together quantum-mechanically.  In quantum mechanics,
states can mix only if they have the same values of conserved quantum
numbers.  We must watch out for this restriction whenever we try to give mass to a
massless particle.

We can understand the required mixing in a very explicit
way for  particles of spin~$\half$.  A massive spin-$\half$
particle is described by the Dirac equation.   The Dirac field is a
four-component field.  In the basis in which the chirality operator
$\gamma^5$ is diagonal, the spinors decompose as 
\beq
          \Psi = \pmatrix{ \psi_L \cr \psi_R}  \qquad\mbox{with} \qquad \gamma^\mu =
          \pmatrix{ 0 & \sigma^\mu\cr \bar\sigma^\mu & 0\cr}\ , \quad  \gamma^5 =
          \pmatrix{ -1 &0 \cr 0 & 1\cr }
\eeq{psidecomp}
In this notation, the familar Lagrangian for the Dirac equation 
\beq
      \La = \bar\Psi ( i \dslash - m) \Psi  
\eeq{firstDirac}
takes the form
\beq
     \La = \psi_L^\dagger i \bar\sigma^\mu \del_\mu \psi_L 
     + \psi_R^\dagger i \sigma^\mu \del_\mu \psi_R  - m
     (\psi^\dagger_L \psi_R + \psi^\dagger_R \psi_L) \ . 
\eeq{Diracdecomp}
For $m = 0$, this formula makes explicit 
that the massless Lagrangian contains the
left-handed ($J = -\half$) and right-handed ($J = + \half$) fermions
completely separately.  This remains true when we couple to gauge
fields, since this coupling is done by replacing the derivative by the
covariant derivative
\beq
           \del_\mu \to D_\mu = (\del_\mu -  i \sum_a g_a A^a_\mu)  \ ,
\eeq{covariantD}
which acts separately on $\psi_L$ and $\psi_R$.   The fields $\psi_L$
and $\psi_R$ are said to have definite {\it chirality}.
The mass term in \leqn{Diracdecomp} mixes these definite chirality
states
to form the four states of a massive spin-$\half$ particle and its antiparticle.
In order for this mixing to be permitted, these two fields must have the
same quantum numbers~\cite{Majorana}.

However, we know that the left- and right-handed quarks and leptons do
not have the same quantum numbers.  The essence of parity violation in
the weak interactions is that the $W$ bosons couple to the  left-handed
chirality states and not to the right-handed chirality states.  In the
SM, the $W$ and $Z$ bosons are gauge fields.  Then the
couplings of each quark or lepton reflect its quantum numbers under
the $SU(2)\times U(1)$ gauge group.  The left-handed fields have $I = \half$ under
$SU(2)$ while the right-handed fields have
$I = 0$, and the two fields also differ in their values of the
hypercharge $Y$. 

We can escape this problem if the $W$ and $Z$ bosons are not gauge
fields.
However, the evidence for gauge invariance in the weak interactions is overwhelming.
Gauge invariance implies the equality of the $SU(2)\times U(1)$
couplings $g$ and $g'$ in the couplings of each species of quarks and
leptons.  Thus, the quality of the overall precision electroweak
fit~\cite{LEPEW,LEPEWtoo} is a consequence of gauge invariance.   The
Yang-Mills structure of the $SU(2)$ vector boson interactions
specifies the structure of the $WW\gamma$ and $WWZ$ couplings, which
are measured to be in agreement with this prediction to percent
accuracy~\cite{triple}.

The only other way to escape this problem is to assume that
$SU(2)\times U(1)$ is a symmetry of the $W$ and $Z$ boson equations of
motion that is not respected by the ground state of the electroweak
interactions.  That is, the symmetry must be spontaneously broken.
If this is true, there must be an agent that is responsible for the
symmetry breaking.   This would be an elementary or composite
field that transforms under the $SU(2)\times U(1)$ gauge group and 
nevertheless acquires a nonzero value throughout space.
To preserve translation and Lorentz invariance, the field must be a
scalar with a constant value, independent of position.
Quarks, leptons, and vector bosons could make symmetry-violating
transitions through their coupling to this nonzero field value.

For electroweak spontaneous symmetry breaking (henceforth, EWSB), 
we call this field the {\it Higgs field}.
Any such field must have quantum excitations, corresponding to
perturbations that change the vacuum value of the field as a function of space and
time. The lowest mass excitation is called  the {\it Higgs boson}.

\subsection{The Higgs mechanism}

Generating mass for spin-1 particles such as the $W$ and $Z$ bosons
brings in additional difficulties.   Even before we add the mass term,
the reduction of a 4-vector field $A_\mu(x)$ to the Poincar\'e
representation with helicity 1 and $-1$  states is quite delicate.
The simplest idea is to quantize the vector field as one would a
scalar field.   This leads to a set of creation and annihilation
operators with the Lorentz-invariant commutation relations
\beq
      [  a^\mu_p, a^{\nu\dagger}_k]   =  ( -g^{\mu\nu}) (2\pi)^3
      \delta(\vec p- \vec k) \ .
\eeq{arelation}
The metric has $(-g^{00}) = -1$, and so the operator $a^{0\dagger}_k$
creates states with negative norm, implying negative probability.   The
interactions of the theory must be set up so that these negative
probability particles can never be emitted.  Quantum Electrodynamics,
and, more generally, Yang-Mills gauge theories, achieve this.   The
proof makes strong use of local gauge 
invariance~\cite{unitarityone,unitaritytwo,unitaritythree}.

To give mass to a spin-1 particle, we must  add to the helicity $\pm1$
states a helicity 0
state that will complete the needed complement of 3 = $(2J+1)$ spin
states.
We must do this  without disrupting the cancellation of negative probability
states described in the previous paragraph. 

The achievement of Higgs, Brout and Englert, and Guralnik, Hagen, and
Kibble~\cite{Higgs,Brout,Guralnik} was to show that spontaneous breaking of the gauge
symmetry achieves this in a Lorentz-invariant way.
Spontaneous breaking of a continuous symmetry automatically generates
a massless particle, the Goldstone boson~\cite{Goldstone}.   This
boson is created and destroyed by the current associated with the
symmetry. In a gauge theory, this same current defines the coupling of
the vector boson.  So, the vector boson and the Goldstone boson
naturally mix.  The fact that this mixing preserves the gauge
invariance of the equations of motion and the natural cancellation of
negative probability states was already known from studies of
superconductivity~\cite{Nambu,Anderson}.    It is a nice bonus that
the mixing is also fully relativistic.

\begin{figure}
\begin{center}
\includegraphics[width=0.70\hsize]{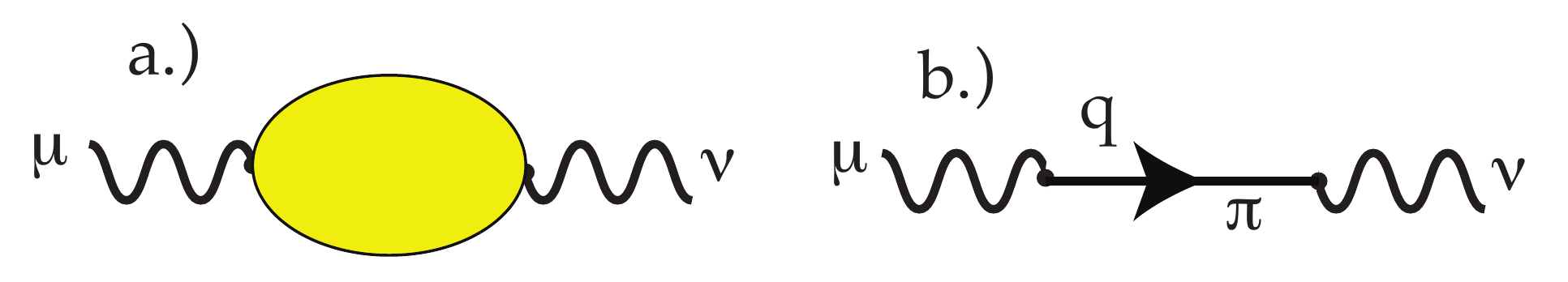}
\end{center}
\caption{(a) Vector boson self-energy diagram; (b) contribution to
  this diagram from a Goldstone boson intermediate state.}
\label{fig:Higgsmech}
\end{figure}

It is not so difficult to describe the process of mass generation
more explicitly.  At zeroth order, a gauge field  obeys Maxwell's
equations and thus must be massless.  A mass can potentially be
generated by the self-energy diagram shown
in Fig.~\ref{fig:Higgsmech}(a).  For a scalar field, the analogous
diagram would directly shift the mass to a nonzero value.   For a gauge field, however, the
currents must be conserved.   The diagram must, then, have the form
\beq
        -i g^2 \biggl( g^{\mu\nu} -{ q^\mu q^\nu\over q^2}\biggr) \Pi(q^2) \ .
\eeq{Pirel}
The vector boson obtains a mass if 
\beq
      \Pi(q^2)  \neq  0  \  \mbox{at} \ q^2 = 0 \ .
\eeq{MfromPi}
However, 
if all states created by the current are massive, this diagram cannot
have a singularity at $q^2 = 0$.   We must find
   $   \Pi(q^2) \sim q^2 $, and so the vector boson mass cannot be
   moved away from  zero.   This is the
   standard argument in QED  that quantum corrections do not generate a mass
   for the photon.

Spontaneous symmetry breaking alters this conclusion.  In a
system with spontaneous breaking of a continuous symmetry, there is a
Goldstone boson. The Goldstone boson is created and destroyed by the 
current that couples to the gauge field.  This is the same current
whose 2-point function gives the vacuum polarization diagram in 
\leqn{Pirel}.   The matrix element for the current to destroy the
Goldstone 
boson is written
\beq
      \bra{\pi(q)}  j^\mu(x) \ket{0} = i F q^\mu \ ; 
\eeq{createG}
Lorentz invariance fixes the form of the matrix element, requiring a 
 parameter $F$ with the dimensions of mass. From the form of 
\leqn{createG}, we find a contribution to the 
vector boson self-energy shown
in   Fig.~\ref{fig:Higgsmech}(b),
\beq
       (-i gFq^\mu) {i\over q^2} (igF q^\nu)  \ .
\eeq{Gcongtrib}
This is  compatible with \leqn{Pirel}, the complete self-energy, only
if that expression reduces,
as $q^2 \to 0$,  to 
\beq
    - i (gF)^2 \biggl(g^{\mu\nu} -{ q^\mu q^\nu\over q^2}\biggr) \ .
\eeq{Pireltwo}
We see that the  presence of the Goldstone boson requires a mass for the vector 
boson
\beq
     m^2 = (gF)^2 \ .
\eeq{getamass}
This is the Higgs mechanism.

\subsection{Goldstone boson equivalence}

There are many formal questions that one might ask about the simple
argument given in the previous section.  Of these, there is one that
is particularly important to discuss:  What is the eventual spectrum
of particles left by the mechanism?  Does it successfully remove all
massless states, and all negative-metric states?   The short answer to
this question is that the combination of the vector field and the
Goldstone boson field leads to precisely three physical states, the
three polarization states of a massive vector boson.  Depending on the
gauge chosen to represent the gauge field, there may be additional
unphysical states, of zero or nonzero mass.  However, these states are
merely an artifact of the method of calculation.  They make no
contribution to the scattering amplitudes of the physical vector boson
states and associated matter particles.  For a careful proof of this
statement, see \cite{unitarityone,unitaritytwo,unitaritythree}.

However, the Goldstone boson does not quite disappear. There are
definite phenomenological consequences of its role in the vector boson
mass generation.    Recall that the role of the Goldstone boson is to
provide a helicity 0 state that combines with the helicity $\pm 1$
gauge boson states to provide the 3 states required for a massive spin
1 particle.  If the final massive boson is near rest, its states will
be a mixture of gauge boson and Goldstone boson states that might be
difficult to untangle.  However, for a highly boosted massive vector
boson, there should be a clear distinction between the helicity $\pm 1$
states, which should have properties similar to massless transversely
polarized 
gauge bosons,
and the helicity 0 state, which should have properties similar to
those of the original Goldstone boson.

This intuition is made precise by the Goldstone Boson Equivalence
Theorem (GBET) of Cornwall, Levin, Tiktopoulos, and
Vayonakis~\cite{Cornwall,Vayonakis}.
The theorem states that the amplitude for producing the helicity 0
state of a massive gauge boson, considered in the limit of large
momentum, becomes equal to the amplitude for producing the original
Goldstone boson.  For example, for the $W$ boson of the SM,
\beq
   \M( X \to Y + W^+_0) \approx \M(X \to Y + \pi^+) + {\cal
     O } ({m_W^2\over E_W^2} ) \ .
\eeq{GBET}
where $\pi^+$ is the Goldstone boson associated with the spontaneous
breaking of the $SU(2)$ gauge symmetry.
Chanowitz and Galliard have given a 
 careful statement and proof of this theorem, valid for any number of
vector bosons~\cite{Gaillard}.

A surprising consequence of the GBET is seen in top quark decay.  The
top quark decays by the simple process $t\to W^+ b$, to an on-shell
$W$ boson.
Naively, one might expect that the rate of this decay would be an
expression of the form 
$\Gamma_t \sim g^2 m_t$, where $g$ is the weak interaction
coupling constant.   Really, though, the formula for the 
decay rate is
\beq
\Gamma_t = {g^2m_t\over 32\pi}\ ({m_t^2\over 2m_W^2} +1)\  ( 1 - {m_W^2\over m_t^2})^2 \ ,
\eeq{topdecay}
ignoring the bottom quark mass and higher-order quantum corrections.
One term in the formula for the rate is enhanced by a factor $m_t^2/2m_W^2$. 

The formula \leqn{topdecay} can be derived without any reference to
gauge invariance.  One simply uses the standard form of the V--A weak
interaction coupling
\beq
         \Delta\La =  {g\over \sqrt{2}} W_\mu \ \bar b_L \gamma^\mu t_L  
\eeq{weakcoupl}
and sums over the three polarization states of the massive $W$
boson in the final state.    The enhancement  can be seen to be
associated with the decay  to a helicity 0 $W$ boson.
 However, the result only makes sense when
viewed from the perspective of the GBET.  According to the GBET, the
top quark couples to the helicity 0 component of the $W$  boson with
the strength of the top quark coupling to the Higgs field, which is larger than the
coupling of the $t$ to the weak interactions.  More explicitly, if $v$
is the vacuum value of the HIggs field,
these couplings have the ratio
\beq
  {  y_t^2 \over  g^2 } = {y_t^2 v^2/2\over g^2
   v^2/2} =  {m_t^2\over 2 m_W^2}
\eeq{whatisy}
This is exactly the enhancement factor seen in \leqn{topdecay}. 

 A
corollary to this argument is that the $W$ bosons in top decays should
be polarized, with 
\beq
  { \Gamma(t \to b W_0)\over \Gamma(t \to b W_{T})} =  {m_t^2\over 2
  m_W^2} \approx  2.3 \ .
\eeq{toppol}
This prediction is confirmed by measurements at the  Tevatron 
and the LHC~\cite{CDFt,ATLASt,CMSt}.
Apparently, the $W$ boson secretly knows that its mass comes from the
Higgs mechanism even in processes in which the Higgs boson is not
involved directly.

\begin{figure}
\begin{center}
\includegraphics[width=0.95\hsize]{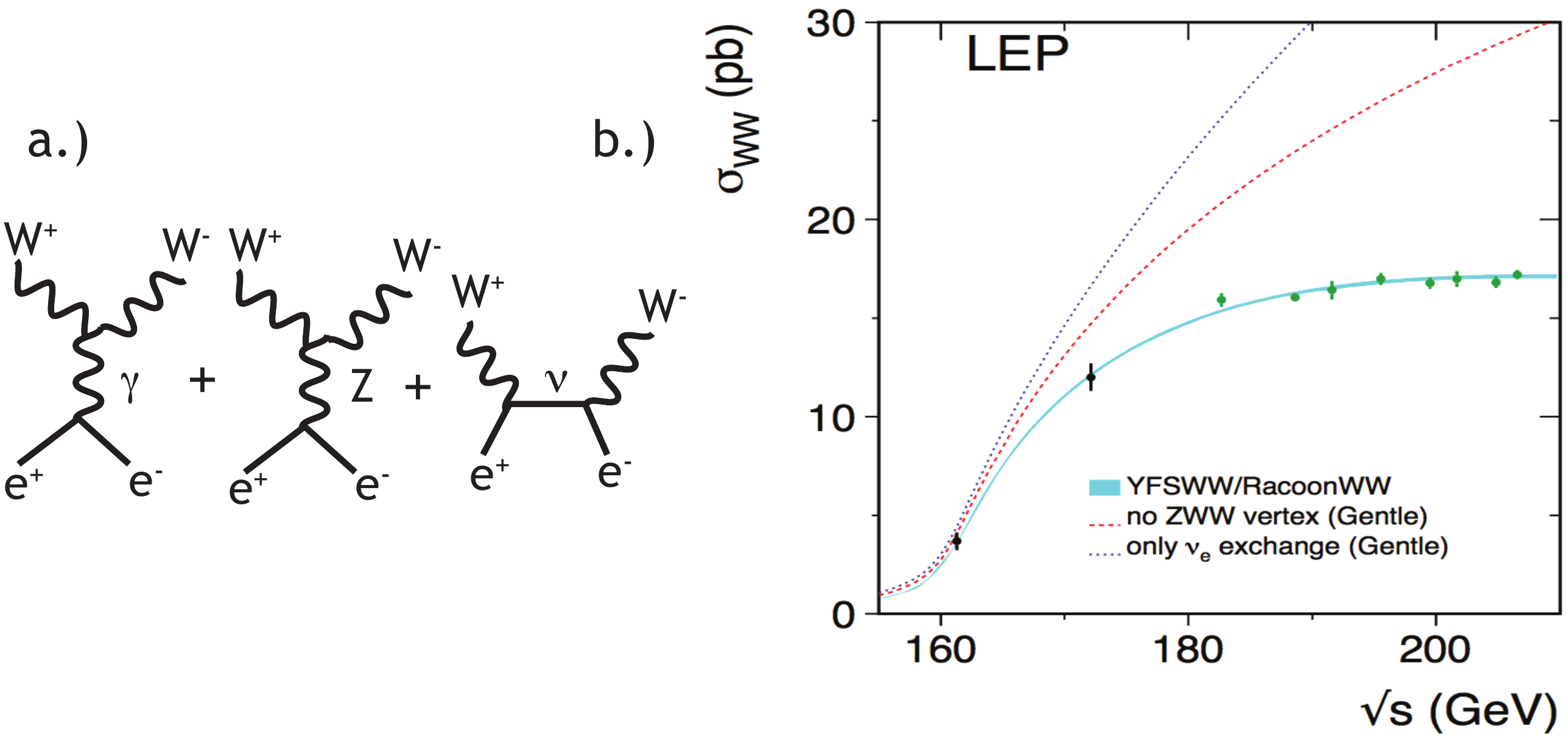}
\end{center}
\caption{(a) Leading-order Feynman diagrams contributing to the cross
  section for $\ee\to W^+W^-$; (b) measured values of this cross
  section from the LEP experiments, from~\cite{triple}.}
\label{fig:WWcross}
\end{figure}

The GBET controls many other features of the high-energy dynamics of $W$
and $Z$ bosons.  One of the most striking predictions occurs in the
cross section for $\ee\to W^+W^-$ at high energy.   At leading order,
there are three Feynman diagrams that contribute to this process,
shown in Fig~\ref{fig:WWcross}(a).  Figure~\ref{fig:WWcross}(b) shows the
measurement of this cross section by the LEP experiments, together
with the prediction of the SM~\cite{triple}.   The figure also shows the
effect of omitting the $Z$ and photon diagrams. The individual diagrams
have a different energy-dependence than the final answer, growing more
strongly with energy by a factor of $s/m_W^2$. Remarkably, the $WW\gamma$ and
$WWZ$ vertices are correctly structured to give an almost complete
cancellation of the three diagrams at high energy~\cite{Buras}.  This
cancellation is difficult to understand from the point of view of the
diagrams in Fig.~\ref{fig:WWcross}(a).  It makes more sense from the
point of view of Goldstone boson equivalence.  The cross section
for the production of two longitudinally polarized $W$ bosons, $\ee
\to 
W^+_0 W^-_0 $ can equal the cross section for the production of a pair
of Goldstone bosons, $\ee \to   \pi^+\pi^-$, only if the terms with
an extra power of $s/m_W^2$ cancel in the full calculation.

Thus, there are many aspects of elementary particle behavior that do
not involve the Higgs boson directly but nevertheless require the
Higgs mechanism for their explanation.

\section{The Standard Model of the Higgs field}

We have now seen that the Higgs mechanism and EWSB  are essential
parts of the gauge theory of weak interactions.  How can this symmetry breaking be
implemented?  The simplest choice is to have one $SU(2)$ doublet of
scalar fields, with a potential that causes this field to acquire a
vacuum value.  The full structure of the SM is given by this one
doublet scalar field, together with the $SU(3)\times SU(2) \times U(1)$
gauge bosons and the quarks and leptons.    In this section, I will review some properties of
this model as it relates to the Higgs field.

\subsection{Formulation of the model}

\begin{figure}
\begin{center}
\includegraphics[width=0.25\hsize]{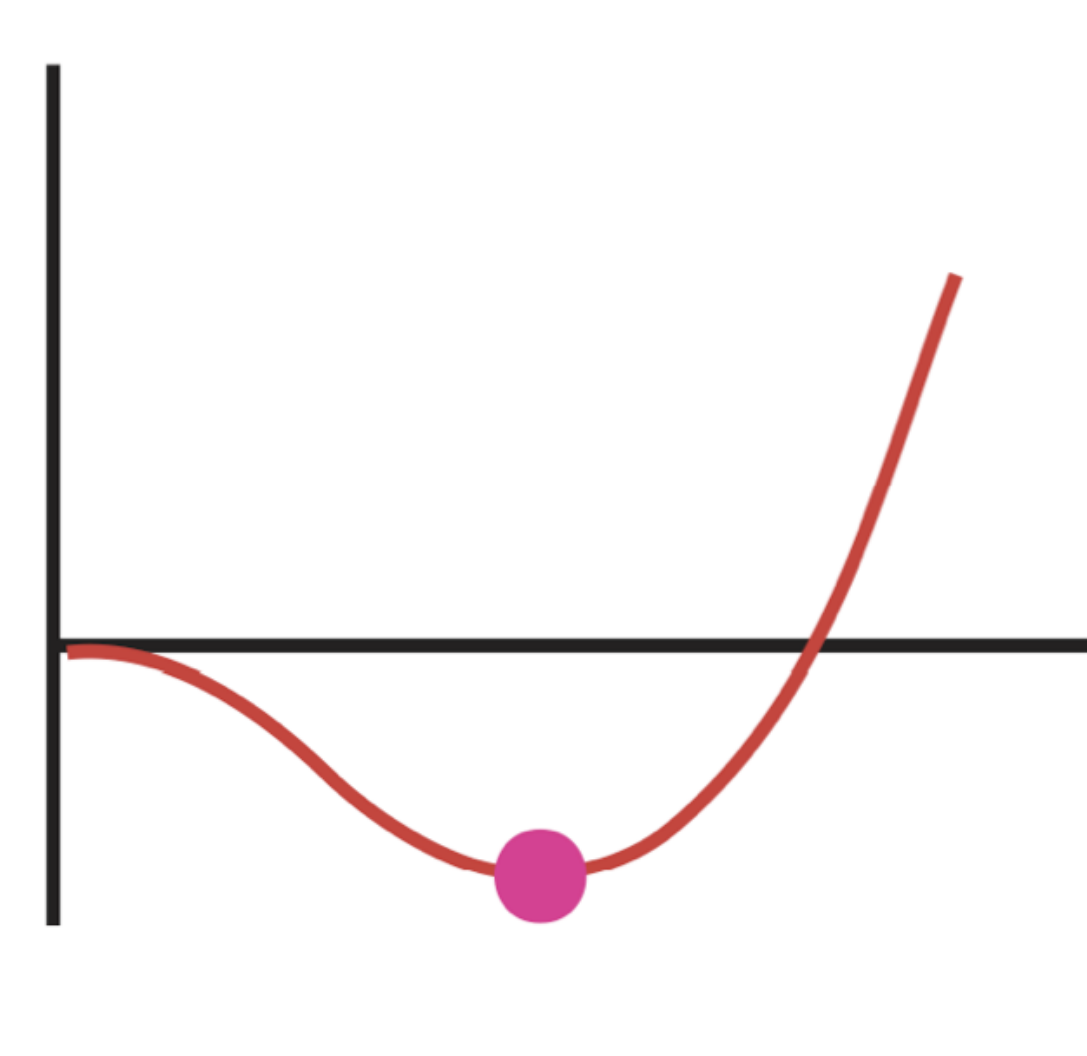}
\end{center}
\caption{Potential energy of the Higgs field in the Standard Model.}
\label{fig:Higgspotential}
\end{figure}

The symmetry-breaking sector of the SM is built from a 
scalar field $\varphi_a(x)$, $a = 1,2$,  that transforms under $SU(2)\times U(1)$
with $I = \half$ and $Y = \half$.  The renormalizable Lagrangian for
this field takes the form 
\beq
 \La  =   |D_\mu \varphi|^2 - V(\varphi) 
\eeq{LagSM}
with 
\beq 
  V(\varphi) =  \mu^2 |\varphi|^2 + \lambda |\varphi|^4 \ .
\eeq{VSM}
The choice 
\beq
     \mu^2 < 0
\eeq{mulez}
leads to a potential of the form shown in Fig.~\ref{fig:Higgspotential}.   The
minimum occurs at 
\beq
    \VEV{\phi} = {1\over\sqrt{2}} \pmatrix{0 \cr v\cr} , \quad \mbox{with} \quad
   v =   \sqrt{-\mu^2/\lambda} \ .
\eeq{vdefin}
From the relation between $v$ and the $W$ boson mass
\beq
     m_W =  g v/2 
\eeq{mWeq}
and our excellent knowledge of the value of the $SU(2)$ coupling $g$
from precision electroweak measurements, we have 
\beq
   v = 246\ \mbox{GeV} \ .
\eeq{valofv}
The related parameters $v$ and $\mu^2$ are the only dimensionful
parameters in the SM.  So, within the SM, \leqn{valofv} sets the scale
for all quark, lepton, and vector boson masses.

\subsection{Natural flavor conservation}

In the SM, the masses of vector bosons are generated by the Higgs
mechanism, as I have already described.   The masses of quarks and
leptons are generated in a more {\it ad hoc} way.  The SM simply
postulates a coupling of the fermion and Higgs multiplets that
respects the $SU(2)\times U(1)$ gauge symmetry.   Then, when the Higgs
field acquires its vacuum value, the fermions receive mass. 

A remarkable feature of the SM is that no special structure is needed
to produce mass terms of a very simple form. The most general
Higgs-fermion couplings, or {\it Yukawa couplings}, consistent with
renormalizability and 
$SU(2)\times U(1)$ are
\beq
  \Delta\La = Y_e^{ij} \bar L^i \cdot \varphi e_R^j\ +\ Y_d^{ij} \bar Q^i
  \cdot d_R^j \ +\  Y_u^{ij} \bar Q^i * \varphi^*  u_R^j\  +\  h.c.
\eeq{genYukawa}
where $i,j = 1,2,3$ run over the quark and lepton generations, $L^i$,
$Q^i$ are the left-handed lepton and quark fields, and the
contractions of these $SU(2)$ doublet fields with the Higgs field are
\beq
    \bar L \cdot \varphi = \bar L_a \varphi_a \ , \qquad   \bar Q \cdot \varphi = \bar Q_a \varphi_a \ , \qquad \bar Q * \varphi^*
     = \eps_{ab} \bar Q_a \varphi_b^*  \ .
\eeq{contractHiggs}
The matrices $Y_f^{ij}$ may be arbitrary $3\times 3$ complex
matrices.  Thus, apparently, they permit arbitrarily strong flavor and
CP mixing.

This mixing, however, can be removed by simple field redefinitions.
Any $3\times 3$ complex matrix can be represented as
\beq
     Y_f  =   V_{fL}^\dagger \cdot  y_f \cdot  V_{fR}
\eeq{decomp}
where $y_f$ is real, non-negative, and diagonal and $V_{fL}$ and $V_{fR}$ are
independent unitary matrices.  Using this decomposition of the Yukawa
matrices, define
\beq
    e_R^{\prime j} = V_{eR}^{jk} e_R^k\ ,\qquad  d_R^{\prime j} = V_{dR}^{jk}
    d_R^k \ , \qquad
u_R^{\prime j} = V_{uR}^{jk} u_R^k\ .
\eeq{newedu}
and 
\beq
    L^{\prime j} = V_{eL}^{jk} L_R^k\ , \qquad  d_L^{\prime j} = V_{dL}^{jk}
    d_L^k \ , \qquad
u_L^{\prime j} = V_{uL}^{jk} u_L^k\ .
\eeq{newLedu}
This change of variables removes factors of the $V_{f\, L,R}$ from the
Yukawa terms, but it potentially reintroduces these factors into the
kinetic terms and gauge coupling terms of the fermion fields.
However, it may be shown that all factors of the $V_{f\, L,R}$  that might appear
in these terms cancel out, except for a modification of the $W$ boson
coupling to quarks,
\beq
   {g\over \sqrt{2}} W_\mu^+   \bar u_L \gamma^\mu d_L \ \to \
 {g\over \sqrt{2}} W_\mu^+   \bar u_L \gamma^\mu\,  V_{CKM} \, d_L \ ,
\eeq{introdCKM}
where $V_{CKM} = V_{uL} V_{dL}^\dagger$.   The Yukawa couplings of the
Higgs field become completely diagonal,
\beq
       \Delta L =  {y_{ei} h \over \sqrt{2}} \bar e^i_L e^i_R \ +\
       {y_{di} h \over \sqrt{2}} \bar d^i_L d^i_R\  +\ 
       {y_{ui} h \over \sqrt{2}} \bar u^i_L u^i_R\ +\  h.c.
\eeq{diagHiggs}
In the full Yukawa Lagrangian, the matrices $V_{f\, L,R}$  appear only in the
combination $V_{CKM}$.

Finally, the flavor and CP mixings in the original Yukawa term
\leqn{genYukawa} remain only in three physical couplings.  First, they
appear in $V_{CKM}$.  This unitary matrix, the {\it
  Cabibbo-Kobayashi-Maskawa matrix}, is required as the source for all
observed flavor mixing and CP violation in the weak
interactions~\cite{Bfactory}.   Second, the matrix $V_{eL}$ appears in
the neutrino mass matrix, another place where there is known flavor
mixing and the possibility of CP violation~\cite{numass}.  These two 
residual appearances should be considered positive features of the model,
giving flavor and CP violation precisely in the ways that it is
observed in experiments.  Finally,
the sum of the overall phases of the $V_{f\, L,R}$  matrices of quarks shifts the
QCD $\theta$ parameter.   If this parameter  is nonzero, it leads
to $P$ and $T$ violation in the strong interactions, in particular, to
a nonzero
neutron electric dipole moment~\cite{neutronEDM}.   This is a problem
for the SM whose solution requires the introduction of a new particle,
the {\it axion}, or additional symmetries~\cite{strongCP}. 

It is fair to consider that the successes of this analysis
outweigh its problems.   The outcome is referred to as {\it natural
  flavor and CP conservation} in the Higgs interactions.   It is
important to note that this property is typically lost in models of
the Higgs sector that generalize the one in the SM.

\subsection{The end of the universe}

The Higgs sector of the SM has one more, quite unexpected, property.
If the SM is exact up to energies much higher than those currently
probed by accelerators, we can extrapolate its behavior using the
renormalization group.   The most important effect of this is that the Higgs
field self-coupling, which determines the form of the Higgs potential,
evolves with energy scale.  

The dominant terms in the renormalization group equation for the Higgs
self-coupling are
\beq
{d\lambda\over d \log Q} = {3\over 2\pi^2} \biggl[ \lambda^2 + \half
\lambda y_t^2 - {1\over 4}
{y_t^4} + \cdots\biggr]
\eeq{RGforlambda}
where $\lambda$ is the Higgs field self-coupling and $y_t$ is the top
quark Yukawa coupling.   The correction proportional to $y_t^4$ turns
out to be the largest effect
and fixes the sign of the right-hand side.
Then $\lambda$  becomes smaller at higher
energy scales, eventually becoming negative.   With the  current  value of
the top quark mass, $m_t = 173$~GeV~\cite{topquarkmass}, this analysis  predicts
that $\lambda$ will become negative at about $10^{11}$~GeV.   This
means that, if the SM is exact at energies beyond $10^{11}$~GeV, the
conventional vacuum of this model is unstable.  With a very long
half-life,
estimated to be $10^{600}$ yr, the vacuum expectation value of the
Higgs field should tunnel to a very large value, near the Planck scale~\cite{Strumia}.

It is not clear that this instability shoud  be considered
 a problem with the SM.  There is a point of view that 
our universe must be unstable, to avoid the prediction that most intelligent beings in
the universe are 
``Boltzmann brains'', isolated conscious entities produced
spontaneously by quantum processes~\cite{Bousso}.  Then the prediction
of instability would be a positive  feature of the SM.   It is also
possible that the instability might disappear after a plausible
revision of experimentally determined parameters.  Within the current uncertainties,
the top quark mass might well have 
the lighter value, 171.1~GeV, that brings the Higgs potential just
to zero at the Planck scale.   This conjecture is the basis for the
idea of asymptotically vanishing Higgs interaction and the Higgs field
as
the inflaton~\cite{Shaposhnikov}.

\section{What is wrong with the Standard Model?}

So, couldn't the Standard Model with one elementary Higgs field describe everything?  I have
already pointed out the existence of phenomena such as dark matter and
dark energy whose explanation certainly lies outside the Standard
Model.  The model must also be extended with effective operators
to generate  neutrino masses.  But one could pose the question more narrowly:   Couldn't the
Standard Model provide a complete explanation for the phenomena of
elementary particle physics up to the currently conceived limits of
accelerator energies?

As far as accelerator-based experiments are  concerned---aside from some much-discussed
discrepancies such as the value of the muon $(g-2)$~\cite{Muong}---the
SM does an excellent job of explaining the wide variety of elementary
particle phenomena.  The main objection to the idea that the SM is a
complete explanation comes from theory.

\subsection{``Naturalness''}

The SM is a compact description of elementary forces, but,
still, it contains a large number of parameters.   These include the
$SU(3)\times SU(2)\times U(1)$ gauge couplings---$g_s$, $g$, and 
$g'$---the quark and lepton masses, the four CKM mixing angles, and two
parameters from the Higgs field potential, or, equivalently, the Higgs
mass and vacuum expectation value.    This is already 18 free
parameters.    A complete specification of the model starting from the most general
renormalizable Lagrangian with the SM gauge symmetry contains
three $\theta$ angles and the fully general
$3\times 3$ complex Yukawa matrices, giving a total of 62 parameters.
If the fermions of the three generations are distinguished in an
underlying theory,  most or all of these parameters would be
physically
significant, although only the first 18 combinations of these parameters are
measureable through Standard Model reactions.

It is the  faith of physicists that, eventually, we will be able to 
 predict the model parameters from an underlying theory.  However, this
 simply cannot be done within the SM.    In the SM,
 the higher-order corrections to all of these parameters are
 infinite and require renormalization.   That is, we can only make
 sense of the model by fixing these parameters {\it a priori}. 

This problem is particularly galling for the parameters of the
potential energy of the Higgs field.   The renormalizability of the
theory requires that the Higgs potential takes the simple form
\beq
            V(\varphi) =  \mu^2 |\varphi|^2 + \lambda |\varphi|^4 \ ,
\eeq{Higgspotzero}
up to radiative corrections.   For spontaneous symmetry breaking, the
renormalized value of the parameter $\mu^2$ should be negative.  But
not even the qualitative prediction that the symmetry is
broken is a prediction of the model.  The $\mu^2$ parameter could have
either sign; there is no logic that prefers one sign to the other.
Though the potential \leqn{Higgspotzero} is relatively simple, we have
to take its parameter values as given.  We cannot ask where these
values come from.

Thus, the hypothesis that the SM is the complete
description of elementary particle interactions is a statement about
the ultimate limits of physics explanation.  This hypothesis implies
that we cannot  predict the
Higgs boson mass  and the quark and lepton masses, or even the
qualitative form of the Higgs potential, from deeper principles.  Any
such predictions require a model with more structure than is present
in the SM.  

\begin{figure}
\begin{center}
\includegraphics[width=0.60\hsize]{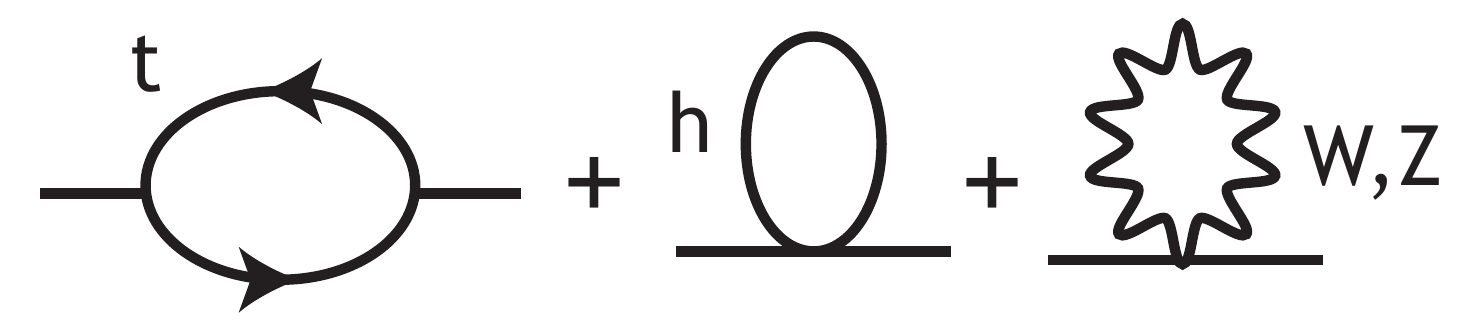}
\end{center}
\caption{One-loop corrections to the Higgs field mass term in the
  Standard Model.}
\label{fig:Higgsmass}
\end{figure}

The form of the first corrections to the $\mu^2$ parameter give a hint
as to where that additional structure might lie.   At the level of
one-loop corrections, the formula for the renormalized $\mu^2$
parameter reads
\beq
 \mu^2 = \mu^2_{bare} - {3y_t^2 \over 8\pi^2} \Lambda^2 + {3\lambda\over 8\pi^2}\Lambda^2  +{ 9 \alpha_w +
3\alpha^\prime\over 16\pi}  \Lambda^2 +
\cdots\ ,
\eeq{muren}
where the corrections written out come from the loop diagrams shown in 
Fig.~\ref{fig:Higgsmass},  with top
quarks, Higgs bosons, and $W$ and $Z$ bosons, respectively~\cite{EJ}.  All of these
diagrams depend quadratically on the ultraviolet cutoff, as shown.
The diagrams have different signs, so even the final sign of
$\mu^2$---which determines whether the symmetry is
broken or not---depends on how these diagrams are regularized.

Setting $\Lambda$ equal to the Planck scale, this equation implies an 
absurd cancellations of the first 30 decimal places of the terms on
the right-hand side of \leqn{muren}   This difficulty
is called the ``gauge hierarchy problem'' and is often presented in
itself as a motivation for new physics~\cite{Dine}.  I have stressed
already that, for me, this
problem is part of a larger problem---that the mass and couplings of the
Higgs boson should be predicted, and that the SM is
inadequate to that task.  

We can, however, use \leqn{muren} as a
rough way to estimate the masses of new particles required in a theory
that predicts the value of $\mu$.   The first thing these particles
must do is to cancel the quadratic divergences in the diagrams of
Fig.~\ref{fig:Higgsmass}.  Let us assume  naively that the calculation of
$\mu^2$ will not entail a cancellation in more than the first decimal
place.  Then we expect new particles of mass less than  2~TeV to
cancel the top quark loop correction, new particles of mass less than
3~TeV to cancel the Higgs loop correction, and new particles of mass
less than 5~TeV to cancel the  $W$ and $Z$ loop corrections.

This is not in itself a proof that new particles must be present in
the energy regime of the LHC and other planned accelerators.  But, it
indicates a tremendous opportunity  for discovery. 

\subsection{Ordering in condensed matter}

Many particle physicists do not consider the spontaneous symmetry
breaking of the weak interaction as a problem in itself.   They feel
that this phenomenon is adequately explained by the Higgs scalar
field, perhaps up to the question raised above of the sign of
$\mu^2$.  

In other areas of physics, there are many examples of spontaneous
symmetry breaking.  I feel that it is important for particle
physicists to make a close study of these systems. They teach us
that the presence of spontaneous symmetry breaking is not a random
choice taken by a physical system but, rather, is always the result of a
comprehensible, and often fascinating, mechanism.   In this section, I
will briefly discuss some examples.

The best understood  example of spontaneous symmetry breaking is that
of superconductivity.  The fact that most metals are
superconducting in their ground states was one of the most puzzling
mysteries of condensed matter physics in the first half of the 20th
century.
The problem was finally solved by Bardeen, Cooper, and
Schrieffer~\cite{BCS}, who noticed that the sharpness of the Fermi
surface at low temperature amplifies the effect of any small
attractive interaction of electrons, such as that due to exchange of
phonons.  This leads to a ground state containing a
thermodynamic number of electron pairs in an ordered condensate.   This model leads to a
successful 
quantitative description of superconductivity~\cite{Tinkham}.

The idea of condensation of fermion pairs stimulated by an attractive
interaction has been applied to other, quite different, physical
systems.  It describes the array of superfluid ground states of
He$^3$~\cite{Hethree}.  Pairing of protons and neutrons is seen in nuclear
spectra (although rigorously, because nuclei have finite size, there
is not true spontaneous symmetry breaking)~\cite{ABohr}.  Nambu and Jona-Lasinio
used a model with pair condensation to explain the spontaneous
breaking of chiral symmetry in the strong interactions~\cite{NJL}.
Unfortunately, in this case, there is no Fermi surface at a nonzero
momentum, so pairing requires that the attractive interaction between
nucleons or quarks be above some critical strength. (If the analogy
with superconductivity were more exact, we would understand the generation of the constituent quark
masses better, and Nambu would have received his Nobel Prize decades
earlier.)   Finally,  the theory of the BCS ground state and its effect on the
electromagnetic field directly stimulated the
original papers on the Higgs mechanism~\cite{Higgs,Brout,Guralnik,Nambu,Anderson}.

Magnets provide other examples of condensed matter sysems with 
spontaneous
symmetry breaking.   Magnetism is confined to a limited
region of the periodic table with almost but not completely filled
$d$ orbitals.  This produces an ordering of electron spins through a
principle called {\it Hund's rule}:  Because the $d$ electrons repel one
another electrostatically, they
tend to favor antisymmetric orbital configurations and therefore
symmetric spin configurations~\cite{Hund}.   When a thermodynamic
number of electrons are involved, the spins of these electrons take on
a common classical
direction, and the ground state violates
rotational symmetry.

Other examples of condensed matter systems with spontaneous symmetry
breaking depend in even more detail on the atomic or molecular forces
involved~\cite{CMP}.  Assemblages of large long or flat molecules lead to the
ordering of liquid crystals~\cite{deGennes}.  Crystal lattices with
soft directions of distortion allow displacive transitions that lower
the crystal symmetry~\cite{MItrans}.    It is wonderful how many
different types of ordering are seen in condensed matter physics and
how, in each case, the nature of the ordering has a direct and
intuitive physical explanation.

Shouldn't this be true also for the spontaneous breaking of the weak
interaction symmetry?  If we fail to search for this explanation, it
will be  an opportunity lost.

\section{Alternatives to the Minimal Standard Model}

What kind of elementary particle theory could provide the explanation
for EWSB?  This question has received a great
deal of thought since Steven Weinberg first discussed it 40 years
ago~\cite{Weinberg}.   The answers that have been proposed would fill
a book.  Here I will briefly introduce the main types of theories
now under consideration.

\subsection{Orientation}

Before I discuss specific models of EWSB, I should emphasize that
these models, though based on very different physical mechanisms,
share
many general  features.  Here  I would
like to highlight three of these.

First,  theories of EWSB  are not simple or minimal in structure.   The reason for
this is that one of the problems a theory of electroweak symmetry
breaking must solve is to render the corrections to the Higgs mass
term shown in Fig.~\ref{fig:Higgsmass}  finite and calculable.  To do
this, some symmetry or other principle must prohibit the Higgs mass
term $\mu^2$ from obtaining divergent corrections from quantum fluctuations of
very high momentum.
This is not straightforward.  The Higgs mass term 
\beq
      \mu^2 |\varphi|^2
\eeq{Hmassterm}
is a scalar under the Lorentz group and under $SU(2)\times U(1)$. It  respects
all other symmetries encountered in a first course on 
quantum field theory.   To forbid this term, we need to invoke more
advanced 
 symmetry principles, for which examples will be given below.  Theories with these structures are not
mere extensions of the SM but have their own profound
implications. 

Second, models of EWSB contain new particles that contribute to
the radiative corrections to the Higgs mass parameter $\mu^2$.   The
higher symmetry of the model might make the Higgs mass parameter
finite in principle, but this mass term must also be finite in
practice when its 1-loop radiative corrections are calculated.  This
requires new particles, with masses suggested to be of TeV size,
 to cancel the divergences due to the heavy SM particles $t$,  $h$,
$W$, and $Z$.

Finally,  many theories of EWSB involve the top
quark in an essential way.  Since the top
quark is the heaviest particle of the SM, and therefore
the one most strongly coupled to the Higgs field, the top quark
contribution in Fig.~\ref{fig:Higgsmass} is the term with the largest
coefficient.  And, this coefficient is negative. Even after  this term is made finite by
adding extra particles in the loop, it is quite plausible that the
result remains negative.
If so, it can be 
the strongest effect driving EWSB.  In the specific models that I 
discuss below, we will see specific physics explanations for why the
contribution of the top quark and its partner particles has the
correct sign to drive EWSB.

\subsection{Supersymmetry}

Although the mass term of a scalar field is not obviously restricted
by symmetry, the mass term of a spinor typically violates some
global symmetry such as a chiral symmetry.   A relation between the Higgs
scalar field and a spinor field then might have the power to prohibit
corrections to the Higgs mass term.  To implement this, we would
postulate a symmetry
\beq 
          \delta \varphi = \bar \eps \, \psi \ ,
\eeq{protoSUSY}
where $\psi$ is a spin-$\half$ field and $\eps$ is a spinor parameter.
A symmetry that connects fields with spin differing by $\half$ unit is
called a {\it supersymmetry}. 

It turns out that the combination of supersymmetry and Lorentz
invariance has very strong implications~\cite{strongSUSY}.  In
theories with 
both symmetries, there must be a conserved spin-$\half$ charge  $Q$ whose square is
the Hamiltonian.  More precisely, the charge  $Q$ satisfies
\beq
                 \{  Q_\alpha,  Q_\beta \} =  2
                 \gamma^\mu_{\alpha\beta}  P_\mu  \ ,
\eeq{totalQ}
where $P_\mu$ is the total energy-momentum of the theory.  Then every
particle in the theory must participate in the supersymmetry.   In a
supersymmetric extension of the SM, not only do the Higgs
bosons have spin-$\half$ partners, but also  the quarks and leptons
have spin-0 partners with the same $SU(2)\times U(1)$ quantum
numbers, and the gauge bosons have spin-$\half$ partners. 

There is a large literature on the spectrum of particles predicted by
supersymmetry and the expectations for the properties of supersymmetric 
particles that might be found at colliders~\cite{Martin,x,y,z}.  In
the remainder of this section, I will focus tightly on the connection
between supersymmetry and  EWSB.

Supersymmetry has a number of specific implications for the Higgs
field.  First, it motivates the presence of scalar fields in the
theory.  In the SM, the Higgs field is the one and only scalar field.
In the supersymmetric extension of the SM, there is a scalar field for each
left- or right-handed fermion.  In addition, the constraints of
supersymmetry imply that it is not possible for a single Higgs field
to give mass to both $u$ and $d$ quarks.  At the minimum, two Higgs fields,
$H_u$ and $H_d$, 
are needed.  $H_d$ can also give mass to the charged leptons.

 The large number of scalar fields brings a new
problem.   The vacuum expectation value of the Higgs field breaks the
weak interaction symmetry $SU(2)\times U(1)$ to electromagnetism,
giving the pattern of symmetry breaking that we see in nature.   But
at first sight, it seems equally possible that one of the other scalar
fields in the theory will obtain a vacuum expectation value.  This
would always be a disaster.  For example, if the scalar partner of
the right-handed top quark were to obtain a vacuum expectation value,
that would leave $SU(2)$ invariant while breaking $U(1)$ and also the
$SU(3)$ color symmetry of QCD.    An explanation for electroweak
symmetry breaking in supersymmetry must also include an explanation of
why the other scalar fields do not acquire vacuum values.

In a theory with exact supersymmetry, the mass parameters for the
scalar fields are highly restricted.  Since the quarks and leptons
cannot have mass terms in the absence of electroweak symmetry
breaking, then also  the associated scalar fields cannot have mass
terms.   The only allowed mass term is one involving the  two Higgs
fields $H_u$ and $H_d$,
\beq
        \mu^2 ( |H_u|^2 + |H_d|^2)
\eeq{muterm}
This is a positive (mass)$^2$.  Though \leqn{muterm}  is renormalized by
corrections that rescale the fields $H_u$ and $H_d$,  there are no additive
 radiative
corrections to this term at any order in perturbation theory.
Essentially, additive corrections from  loop diagrams involving quarks, leptons, and gauge bosons
are cancelled exactly by corrections from similar diagrams involving their  spin 0 and
spin-$\half$ partners.

A realistic model of supersymmetry must be more complex.  Exact
supersymmetry would imply a charged scalar particle with the same mass
as the electron, but no such particle exists.   The most
straightforward way to resolve this problem is to assume that
supersymmetry is spontaneously broken.  It can be shown that
spontaneous supersymmetry breaking among any particles in
nature---even unknown particles with very large masses---will
eventually
feed down to the particles of the SM and produce masses for the
partners of the quarks, leptons, and gauge bosons.   We can then make
a model of EWSB along the following lines:
Spontaneous breaking of supersymmetry gives mass to some new particles
at very short distances, and this in turn gives mass to the
supersymmetric partners of SM particles.
 Radiative corrections involving those mass terms can then
induce  a potential energy function that can favor a nonzero vacuum
value for the Higgs field.

\begin{figure}
\begin{center}
\includegraphics[width=0.30\hsize]{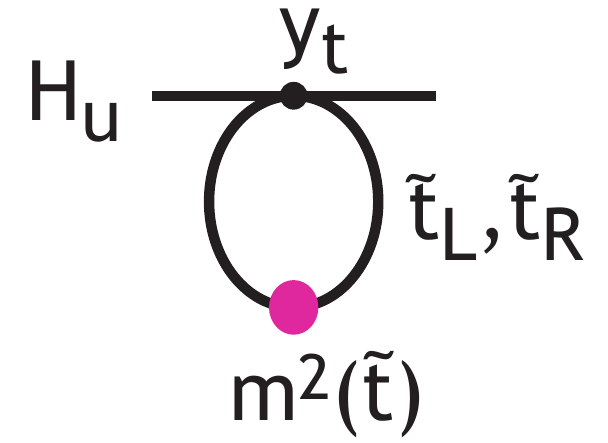}
\end{center}
\caption{A diagram by which the supersymmetry-breaking mass terms of
  $\s t_L$, $\s t_R$ renormalize the mass term of the Higgs field $H_u$.}
\label{fig:oneloopstop}
\end{figure}

Specifically, this can work as follows:   Three scalar fields are
coupled by the top quark Yukawa coupling---the scalar partners of the
left- and right-handed top quarks, $\s t_L$ and $\s t_R$,  and the
Higgs field $H_u$.  All three fields receive mass from supersymmetry
breaking.   Arrange that these (mass)$^2$ terms are all positive,
approximatey equal, and of
TeV size.  Then compute the 1-loop corrections to these mass terms,
which come from diagrams of the form of Fig.~\ref{fig:oneloopstop}. 
This correction is negative by explicit calculation~\cite{Ibanez,Wise,Inoue}.  All three
(mass)$^2$ terms
receive these negative contributions, but the correction to the Higgs
mass is largest, because of the factor of 3 from QCD color flowing
around the loop.  (The $\s t_L$ and $\s t_R$ mass terms also receive positive
corrections from diagrams involving the supersymmetric partner of 
the gluon.)  This calculation creates a potential energy function
with a negative (mass)$^2$ for the $H_u$.  It explains why this scalar
field---and no other---obtains a vacuum expectation value.

Ultimately, this model of electroweak symmetry breaking is testable.
The masses of the top quark partners and the Higgs boson spin-$\half$
partners should not be too far above the 1 TeV mass scale.  The Higgs
partners, which are very difficult to discover at the LHC, could still
be as light as 100~GeV~\cite{HiggsatLHC,HiggsatLHCtwo}. 
  If we could discover these particles and
measure their masses and decay products, it will be possible to
extract all of the parameters that enter the calculation of the Higgs
potential~\cite{computeHiggs}.
If all of the pieces fit together, we could then claim to understand
EWSB at the same level at which we understand
the appearance of superconductivity in metals.

\subsection{Higgs as a gauge boson}

Supersymmetry produces a calculable theory of  the Higgs mass term in
the following way:  At the level at which the symmetry is exact, this
symmetry
strongly constrains  the Higgs potential.  If the symmetry is then softly broken, new terms appear
as  calculable radiative corrections.  These latter terms give the
negative (mass)$^2$ that drives electroweak symmetry breaking.  Other
types of models can implement this same philosophy in different ways.

Another familiar principle that can keep a particle mass at zero is local
gauge invariance.   We can use this in the Higgs story through the
idea of {\it gauge-Higgs unification}~\cite{gaugeHiggsone,gaugeHiggstwo}.   In
this approach, we assume that the universe is 5-dimensional, with the
5th dimension compact and small.  A gauge field $A_M$ in 5 dimensions
has 5 components, $M = 0,1,2,3,5$.  In the compactified geometry, the
first four components make up a 4-dimensional gauge field.  The last
component $A_5$ transforms as a scalar field in 4 dimensions.  If the
full gauge group of the 5-dimensional theory is larger than
$SU(2)\times U(1)$ and contains  fields that transform like the
doublet of $SU(2)\times U(1)$, we can interpret the $A_5$ component of
those extra gauge fields as a Higgs boson multiplet. 

 A simple example
is given by assuming the gauge group of the 5-dimensional theory to be
$SU(3)$.  Let  $t^A$ be  the $3\times 3$ matrices that generate
$SU(3)$.  
The  gauge fields of $SU(3)$ in 5~dimensions take the form
\beq
       A_M^A t^A =  \pmatrix{  A^a_M  \sigma^a/2  + \half B_M  &   \Phi_M\cr
         \Phi^\dagger_M    &   -  B_M  \cr }
\eeq{SUthree}
where $\sigma^a$ are the usual $2\times 2$ Pauli sigma matrices and
$\Phi_M$ is $2\times 1$.

The $SU(3)$ symmetry can be broken by boundary conditions in the 5th dimension.
After appropriate symmetry breaking, the fields $A_\mu^a$ will be the
gauge fields of the weak interaction $SU(2)$,  and $B_\mu$ will combine
with another $U(1)$ to provide the weak interaction $U(1)$ gauge
field. The components $\Phi_M$ have the quantum numbers  $(I,Y) =
(\half, \half)$, and so the doublet $\Phi_5$ has just the quantum
numbers of the Higgs scalar doublet.   Similar constructions can be
made using larger gauge groups for the 5-dimensional theory.  An
attractive choice is to take 
the 5-dimensional gauge symmetry to be $SO(5)$.  This group has a subgroup
$SO(4) = SU(2)\times SU(2)$,  such that one can interpret one $SU(2)$
as the weak interaction isospin gauge group.  The
$SO(5)$ bosons not contained in $SO(4)$ have the quantum numbers
of Higgs fields~\cite{Contino}.  A vecuum expectation value of one of
these $SO(5)$ bosons breaks $SU(2)\times SU(2)$ to a diagonal $SU(2)$
group.   That unbroken symmetry can be interpreted as the 
{\it custodial}
symmetry that protects the relation $m_W^2 = m_Z^2 \cos^2\theta_w$
from receiving large radiative corrections~\cite{custodial}.  

Electroweak symmetry breaking will take place if there is an energetic
reason why the 5th component of the 5-dimensional gauge field should
take on  a nonzero value.   In
a 5-dimensional theory with a periodic 5th dimension, there is in fact
a good reason for this,
 the
{\it Hosotani-Toms mechanism}~\cite{Hosotani,Toms}.   If $A_M$ is a
5-dimensional gauge field, a particle travelling around the 5th
dimension acquires a phase
\beq
                 W =    \exp[ i g \oint dx^5   A_5 Q ] 
\eeq{Wilsonphase}
where $Q$ is the charge of the field under the gauge symmetry
associated with $A_M$.  For
bosons, the energy is typically minimized when $W = 1$ and maximized
when $W = -1$.  However, for fermions, it is the reverse: the energy
is minimized for $W = -1$.   A somewhat formal way to understand this
is to recall that the  functional integral representation of the
thermodynamic 
partition function for fermions uses fermion fields that are
antiperiodic around a compact Euclidean
direction. 
Though this mechanism, a $(t,b)$ quark doublet in the 5-dimensional
space can force the quantity
\beq
       W =    \exp[ i g \oint dx^5  \Phi_5 ] 
\eeq{WilsonphasePhi}         
to be nonzero, where $\Phi_M$ is the off-diagonal gauge field
multiplet indicated in \leqn{SUthree}.  From a 4-dimensional point of
view, this is the induction of a negative (mass)$^2$ for the Higgs
field by radiative corrections due to a heavy top quark and associated
heavy quarks.
The same physics appears in other examples of compact 5-dimensional 
geometries.  In particular, in the 
Randall-Sundrum warped 5-dimensional spacetime~\cite{RS}, a
very similar computation shows that radiative corrections due to a
heavy top quark can drive some $A_5$ with the quantum numbers of the
Higgs field  to acquire a 
vacuum expectation value~\cite{Nomura}

The  5-dimensional picture for the creation of the Higgs potential
implies that this potential is free of ultraviolet divergences.  The
reason for this is that the 
phase factor \leqn{Wilsonphase} is nonlocal over the 5th dimension.
Quantum fluctuations smaller than the full size of the 5th dimension
see only a part of the integral in \leqn{Wilsonphase} and cannot
distinguish this from a local gauge transformation.   From a
4-dimensional point of view, though, the elimination of divergences
seems quite surprising. The states of a  $(t,b)$ multiplet in
5~dimensions can be represented in terms of their momentum in
4~dimensions and their momentum around the 5th dimension.   In the
4-dimensional description, the
momenta in the 5th dimension appear as contributions to the masses
of the 4-dimensional particles. Thus, the 5-dimensional $(t,b)$ multiplet is seen in
4-dimensions as a 4-dimensional $(t,b)$ multiplet plus an  infinite
number of more massive states, called {\it Kaluza-Klein states},  with
the same SM quantum numbers as $t$ and $b$. 
Each of these states gives a quadratically divergent contribution to
the Higgs boson (mass)$^2$.  But, by what seems to be   a miracle, the sum of
these  contributions  is finite.

If these 5-dimensional theories are correct, we will discover the
heavy partners of $t$ and $b$ one by one as we search for new
particles at higher energies.  By measuring the properties of these 
particles,  we could in principle extract their couplings to the Higgs field
and directly verify the cancellation of divergences and the generation
of a finite, negative Higgs (mass)$^2$.

\subsection{Higgs as a Goldstone boson}

There is a third way to construct a model in which the Higgs boson begins as a
massless particle and acquires negative (mass)$^2$ by radiative
corrections.  The methods begins by assuming a new set of strong interactions
at an energy scale well above the electroweak scale, say, 10~TeV, that
breaks a global symmetry to a subgroup that contains $SU(2)\times
U(1)$.  This symmetry breaking produces Goldstone bosons, one for
each broken global symmetry direction.   It is easy to arrange that
some of these Goldstone bosons form a multiplet that transforms as 
 $(I,Y) =
(\half, \half)$ under $SU(2)\times U(1)$.  We can identify this
multiplet with the Higgs scalar doublet.  We can now add $SU(2)\times
U(1)$ gauge interactions and other weak couplings. These will lead to 
radiative corrections that will generate a nonzero potential function
for the Goldstone boson fields and drive these fields to acquire
nonzero vacuum expectation values. 
Models of this type are known as {\it Little Higgs}
models~\cite{LHone,LHtwo}.

A theory of Goldstone bosons is described by a Lagrangian that is
invariant under the original global symmetry.  The global symmetry may
be nonlinearly realized, but still there are significant constraints
that come from this structure.  In particular, these models also
require new heavy quarks with charge $+\tthird$, with vectorlike
couplings to the weak interactions.  As we saw in the case of 5-dimensional models,
these heavy quarks can partially cancel the radiative correction to the Higgs
(mass)$^2$ from the top quark, making this correction
ultraviolet-finite but still negative~\cite{LHone}.  The cancellation
of ultraviolet divergences implies relations between  the Higgs
couplings of the new heavy quarks and that of the top quark, and these
could eventually be tested experimentally~\cite{Aaron}. 

There is a connection between Little Higgs models and the 5-dimensional
models discussed in the previous section.
According to the AdS/CFT correspondence of
Maldacena~\cite{Malda,adsreview}, a scale-invariant quantum field
theory model in 4~dimensions has an equivalent representation as a
5-dimensional model in which global symmetries of the first theory
become local gauge symmetries of the second theory.  This idea allows
the
5-dimensional models in warped backgrounds discussed in the previous
section to be reinterpreted as 4-dimensional models. The Kaluza-Klein
states of  the 5-dimensional theory are reinterpreted as the spectrum
of bound states of the strongly coupled 4-dimensional system.  The gauge field
Higgs multiplet in the previous case becomes the Goldstone boson Higgs multiplet
that we have discussed here~\cite{AHPR}.

The three types of models described in this section illustrate in
different ways the possibility of dynamical explanations of the state
of spontaneously broken symmetry required for the $SU(2)\times U(1)$
theory of the weak interactions.  These models are not simple modifications
of the SM.  They require large numbers of new particles
that must eventually be discovered by accelerator experiments at high
energies.

\section{Where is the new physics?}

If the arguments for physics beyond the SM are so compelling, and if
the particle spectra expected are so rich, then why haven't we found
evidence for these particles?   
 I think that every theorist who puts forward arguments similar to those above
is troubled by this question.  In all three models above, the new
particles introduced to explain the Higgs potential would be expected
to have masses at the scale of hundreds of GeV.  For top quark
partners and other particle with QCD interactions, LHC searches
exclude most scenarios with  new colored particles in this mass
range~\cite{exceptions}.  In addition, we have not yet seen signs of
indirect effects of new particles shifting low-energy observables from
their SM values.

Nevertheless, the situation is different now than in earlier eras of
particle physics.  All of the models reviewed in Section 5 have the
property that the new particles that they predict have vectorlike coupling
to the $SU(2)\times U(1)$ gauge group and do not rely on the Higgs
mechanism to obtain mass.   It is likely that any other extension of
the SM that we might consider would also have this property.  The
known particles of the SM fill complete $SU(2)\times U(1)$ multiplets,
leaving no place there for additional chiral particles.  Heavier
chiral fermions are strongly constrained by the cancellation of chiral
anomalies~\cite{anomaly}.
Further, any additional  chiral quarks would have a major effect on
the couplings of the Higgs boson.  A fourth generation of quarks, for
example, would increase the gluon fusion production cross section of
the Higgs boson by a
factor of 9, a result clearly incompatible with the current data~\cite{Vysotsky}.

Quantum field theories with only vectorlike coupling 
obey a general property called the {\it Appelquist-Carrazone Decoupling
Theorem}. This is the statement that, if heavy particles are added to the
theory, any new interactions due to those particles are suppressed by
a factor  $1/M^2$, where $M$ is the mass of the heavy
particle~\cite{ACtheorem}.   For example, if we add a new heavy quark
of mass $M$
to QCD and measure its effects at a scale $Q \ll M$, any new terms are
suppressed by $Q^2/M^2$.  This follows from the fact that the QCD
Lagrangian is already the most general renormalizable Lagrangian one can write that
contains the known quarks and has the QCD gauge symmetry. 
Corrections to the Lagrangian induced by effects of the heavy quark can only change the
quark masses and the QCD coupling, and add higher-dimension operators
with coefficients that explicitly contain $1/M^2$. The shifts of
quark masses and $\alpha_s$ are visible only if we can independently
measure these parameters at energies above the heavy quark mass. If we
cannot, 
 we cannot know that these shifts have
taken place.   Then the only new and observable terms are of order
$1/M^2$. 

For a theory with chiral interactions and spontaneously broken
symmetry, the situation can be quite different.  The strong statement
of decoupling requires that the heavy particles are complete gauge multiplets.
 At previous
stages of our knowledge, our description of particle physics included
some members of $SU(2)\times U(1)$ gauge multiplets but not others,
for example, $s$ but not $c$,  $b$ but not $t$, or the longitudinally polarized
$W$ boson but not the Higgs boson. In this situation,
  corrections involving the  missing states could contribute large terms to
appropriately chosen amplitudes~\cite{precMW}.  Examples are provided by the 
$c$ and $t$ contributions to $K$--$\bar K$ and $B$--$\bar B$
mixing, the top quark loop contribution to precision electroweak
observables such as $m_W$, and the top quark loop contribution to the 
Higgs coupling to two gluons.  We have seen already in Section~2.3 that the
large top quark Yukawa coupling can lead to effects that are much
larger than  expected naively from perturbation theory. But 
now that the full SM particle content has been discovered, we have
returned to the situation in which new particles added to the model
should have vectorlike couplings and virtual effects suppressed as
$1/M^2$. 

Thus, it is quite plausible that new particles outside the SM might be
present in nature but have only minor effects on observations at currently
explored energies.   When we finally reach the new particles
thresholds, we will turn a corner, and a new realm of physics will come
into view.   Large multiplets of new
particles will suddenly appear.  The reality of these particles will
become obvious.  Later, papers will be written about the
surprisingly small effects of  these particles on low-energy
observables.

The most powerful way to search for physics beyond the Standard
Model, now more than ever, is to search for new thresholds at the
highest energy accelerators.  It is exciting that, this year, the LHC
will finally be running close to its design energy.   Over the next
fifteen years, the LHC will open up a territory in which to search for
strongly interacting particles about 3 times greater
than that currently explored, and a territory for particles with only
electroweak interactions---and signatures appropriate to hadron 
colliders---about 4 times larger than the current one

Still, the direct search for new particles at high-energy accelerators
is ultimately limited by the collider energy.   Each step to higher
energy is now a major technical, social, and political  endeavor.   So 
it is important that it is also possible to search for these particles
through new high-precision probes for the $1/M^2$ effects that they induce.

In particular, the next $\ee$ collider will be able to carry out
high-precision studies of the couplings of the Higgs boson and top
quark.
I have emphasized in this review that these two particles stand at the
very center of the mystery of electroweak spontaneous symmetry
breaking.
New particles that would provide an explanation for this question must
necessarily couple to the Higgs boson.   These new particles are also very likely to
couple to or mix with the top quark.  Today, the Higgs boson and the
top quark
are incompletely understood experimentally.  The couplings of
the Higgs boson and the electroweak couplings of the top quark are
measured only to the 20\% level.   The International Linear Collider,
a 500~GeV $\ee$ collider based on superconducting RF cavities, is now
designed and ready for construction.  Experiments at that accelerator
would bring these measurements below the percent level of accuracy and
would be sensitive to the effects of the new particle scenarios
discussed in this review~\cite{ILCTDR,ILCHiggs}.

Above all, we need to keep our faith in the basic tenet of
physics---that the phenomena of nature have explanations, and that
those explanations can be found by probing nature at successively
deeper levels.   New forces and interactions are out there.  In time,
we will find them.

\Acknowledgements

I am grateful to Halina Abramowicz, Allen Caldwell, and Brian Foster
for their invitation to write for this volume, to Raymond Brock and Beate
Heinemann for their encouragement, and to many colleagues at SLAC and
elsewere for discussions of the issues put forward here.
This work was supported by the U.S. Department 
of Energy under contract DE--AC02--76SF00515.

\end{document}